\documentclass[aps,pra,twocolumn,10pt,superscriptaddress,longbibliography, floatfix]{revtex4-1}
\usepackage[utf8]{inputenc}  

\usepackage[T1]{fontenc}     
\usepackage[british]{babel}  
\usepackage[scaled=0.86]{berasans}  
\usepackage[colorlinks=true,allcolors=blue]{hyperref}  
\usepackage{graphicx} 
\usepackage[babel]{microtype}  
\usepackage{amsmath,amssymb,amsthm,bm,amsfonts,mathrsfs,bbm} 
\usepackage{setspace}
\usepackage{braket}
\usepackage{blochsphere}
\usepackage{pgfplots}
\usepackage{float}
\usetikzlibrary{calc,3d,shapes, pgfplots.external, intersections}
\usepackage[bitstream-charter]{mathdesign}
\usepackage{tikz}
\usepackage{mathtools}
\usetikzlibrary{arrows.meta}
\usepackage{csquotes}
\usepackage{multirow}
\usepackage{adjustbox}

\usepackage{xspace}  
\usepackage{pgfplots}
\usepackage{verbatim}
\usepackage[bitstream-charter]{mathdesign}
\usepackage{todonotes}
\usepackage{xfrac}
\usepackage[nice]{nicefrac}
\usepackage{caption}
\usepackage{subcaption}
\usepackage[normalem]{ulem}
\usepackage{array, tabularx, boldline}
\usepackage{cellspace}
\usepackage{makecell}
\usepackage[normalem]{ulem}
\usepackage{color}
\usepackage[mathscr]{euscript}

\makeatletter
    \renewcommand\@make@capt@title[2]{%
     \@ifx@empty\float@link{\@firstofone}{\expandafter\href\expandafter{\float@link}}%
      {\textbf{#1}}\@caption@fignum@sep#2\quad}%
    \makeatother
   
\makeatletter 
\renewcommand{\fnum@figure}{\textbf{Figure~\thefigure}}
\makeatother

\usepackage{accents}
\usepackage{gensymb} 
\usepackage{dcolumn}
\usepackage{bm}
\usepackage{yfonts}
\usepackage{mathtools}
\usepackage{comment}

\usepackage{notes2bib}
\bibnotesetup{
note-name = ,
use-sort-key = false
}

\newcommand\id{\leavevmode\hbox{\small1\kern-3.3pt\normalsize1}}
\newcommand{\ketbra}[2]{\left|#1\middle\rangle\middle\langle#2\right|}

\begin{document}


\title{Daytime and Nighttime QKD over an atmospheric free space channel with passive polarisation bases compensation}

\author{Saumya Ranjan Behera} 
\affiliation{Raman Research Institute, C. V. Raman Avenue, Sadashivanagar, Bengaluru, Karnataka 560080, India}
\author{Melvee George}
\affiliation{Raman Research Institute, C. V. Raman Avenue, Sadashivanagar, Bengaluru, Karnataka 560080, India}
\author{Urbasi Sinha}
\email[]{usinha@rri.res.in}
\affiliation{Raman Research Institute, C. V. Raman Avenue, Sadashivanagar, Bengaluru, Karnataka 560080, India}

\begin{abstract}
\noindent \textbf{Abstract:} Quantum Communication (QC) represents a promising futuristic technology, revolutionizing secure communication. Photon-based Quantum Key Distribution (QKD) is the most widely explored area in QC research, utilizing the polarisation degree of freedom of photons for both fibre and free-space communication. In this work, we investigate and mitigate the challenges posed by fibre birefringence and atmospheric effects on QKD, using a $50$-meter free-space optical link and entanglement-based BBM92 QKD protocol. We implement a passive polarisation correction scheme to address the critical issue of polarisation scrambling induced by fibre birefringence and the difference in the frame of reference between Alice and Bob. This scheme effectively mitigates these adverse effects, ensuring reliable polarisation control over the quantum channel. Furthermore, we conduct QKD experiments in both day and night conditions, encountering challenges such as high background noise levels and dynamic environmental changes. To overcome these issues, we employ various filtering techniques to enhance signal quality and security. Our results demonstrate the successful implementation of QKD over a free-space optical link by producing information-theoretic secure QBER of $<11\%$ on an average and high keyrate, even under varying lighting and weather conditions. Over one 24 hour cycle of data acquisition, we measured an average daylight keyrate and QBER of ($3.9118\pm0.7339 KHz$ and $10.5518\pm1.3428\%$) respectively and night time Keyrate and QBER of ($4.6118\pm0.8088 KHz$ and $10.3545\pm1.2501\%$) respectively. 
\end{abstract}

\maketitle

\section{Introduction}
\label{Introduction}
Quantum Key Distribution (QKD) is a cutting-edge technology that enables secure communication by utilizing the principles of quantum mechanics. It promises unbreakable encryption, making it highly desirable for applications involving sensitive data transmission. The first QKD protocol, proposed in 1984\cite{bennett1984ieee}, is well known as the BB84 protocol. B92 as the first free-space QKD, was demonstrated over a 30cm optical link\cite{B92}. Our experiment employs the BBM92 protocol\cite{bbm92}, a variant of B92 utilizing entangled photons. To achieve secure QKD with photons, quantum channels have to be employed in the form of optical fibre or free-space optical links. In comparison to fibre-based QKD demonstrations, free-space links emerge as an appealing solution for achieving much greater distances, primarily because of their minimal absorption characteristics \cite{bedington2017progress,toyoshima2011polarization}. 
Free-space QKD is important for overcoming the distance limitation of quantum communication systems by incorporating satellite communication. Satellite-based QKD aims at establishing quantum channels at the scale of thousands of kilometres. One of the major practical challenges in satellite-based QKD includes the active control of polarisation fluctuations resulting from photon transmission through the atmosphere or variation of reference frames due to satellite movement.\\

In our present research, we have successfully implemented a polarization correction technique by performing measurements along suitably optimized (rotated) projection bases. This approach holds significant promise for mitigating polarization fluctuations in both free-space and fibre-based QKD systems. Building upon our earlier work\cite{chatterjee2023polarization}, where we initially introduced this novel technique in a controlled laboratory setting and demonstrated its effectiveness, our current experiment takes a step further by demonstrating this over an atmospheric free space link. Our method is designed to achieve a reduced Quantum Bit Error Rate (QBER), which quantifies the rate of errors occurring during the transmission and reception of qubits, as well as an optimized keyrate, representing the rate of successfully transmitted qubits over the quantum channel.\\

Here we demonstrate the feasibility and effectiveness of this approach in a more realistic scenario where a 50-meter-long free-space atmospheric link separates Alice and Bob. Instead of an active feedback mechanism to compensate for the polarisation scrambling, we perform a quantum state tomography to arrive at optimal measurement bases for any one party resulting in maximal (anti-)correlation in measurement outcomes of both parties. Taking into account the polarisation fluctuation in fibre or free-space transmission for QKD, we demonstrate that our choice of optimised measurement bases can retain the QBER below the information-theoretic secure threshold regardless of the fidelity of the measured state with the maximally entangled state. Our ultimate goal is to extend this technology to satellite-based QKD to enhance the security and efficiency of QKD protocols. Furthermore, we also evaluate the feasibility of entanglement-based daylight QKD with our new technique on different days with changing weather conditions as well. We observe the establishment of secure quantum key throughout the day, despite the challenges of daylight operation. \\

The experiment was run continuously over the day from 15th-22nd September 2023, in 24-hour cycles. We show representative data over three such cycles, with different light and weather conditions. In this period we witnessed bright sunlight, rain and partly cloudy weather; over the whole period, the rate of detected pairs and background events varied depending on weather conditions. \\

Secure quantum communications-based approaches are gaining utmost importance in fields such as defence, finance, healthcare, and for government agencies, where the protection of sensitive information is critical. The novelty of our work is to employ the measurement-based approach over a free space optical link to correct polarisation fluctuations, instead of an active feedback mechanism. The approach of using commensurate mutually unbiased measurement bases (MUB) enables more efficient polarisation fluctuation mitigation caused by varying birefringence in free space or fibre media. Our current demonstration over an atmospheric free space link provides a significant advancement over our earlier in-lab demonstration.\\

The significance of our work lies in its contribution to advancing towards \emph{practical} secure quantum communication. The successful demonstration of this technique for QKD over free space opens doors to numerous practical applications. This needs to be followed up by future work where further complexities are introduced in the set-up to enable more and more realistic scenarios. For instance, if this technique could be employed for real-time corrections while the receiver module is moving instead of stationary, that would take us closer to its employability in longer-distance practical QKD, including satellite-based approaches.\\ 

Our manuscript is organised as follows. In section \ref{Introduction} we give a comprehensive introduction to our work and our motivation towards it. Moving on to section \ref{Background} we discuss the theory behind the passive polarisation correction mechanism as well as provide a detailed account of both daylight and nighttime QKD, the challenges involved and a survey of the existing literature. Section \ref{Entangled Source} briefly discusses our entangled photon source. Section \ref{Experiment} is dedicated to an in-depth explanation of our experimental procedures. In section \ref{Results} we present the results of our experiment. Finally, in Section \ref{Discussion}, we conclude our work and discuss its implications for future research activities.

\section{Background}
\label{Background}
\subsection{MUB Approach}    
In a recent work from our group\cite{chatterjee2023polarization}, we devised and demonstrated a technique of choosing the correct measurement bases to counter any change in the polarisation state because of photon propagation in free space or within the fibre. Instead of performing active feedback for polarisation correction if we choose a modified measurement bases based on data obtained from quantum state tomography, we showed how one can obtain high visibility with different bases as well as less QBER and a high keyrate.\\

To implement this correction, we follow the following steps:\\
We start with a density matrix $\rho^{AB}$ obtained through tomographic techniques. The entangled state which we intended to share between Alice and Bob is given by:

\begin{align}
\ket{\psi} = \frac{1}{\sqrt{2}}\left(\ket{H}^A\ket{V}^B + \ket{V}^A\ket{H}^B\right)
\label{Eq:Prepared_state}
\end{align}

From the experimentally obtained density matrix $\rho^{AB}$, we can find the nearest pure state by doing an eigendecomposition such that $\rho^{AB}{=}\sum_i \lambda_i\ketbra{\lambda_i}{\lambda_i}$, where $\{\lambda_{i's}\}$ are the eigenvalues and $\{\ket{\lambda_i}'s\}$ are the corresponding eigenvectors. The eigenvector corresponding to the maximum eigenvalue represents the nearest pure state, as it will have the maximum overlap with $\rho^{AB}$. This can be quantified by defining fidelity as  $F = \bra{\lambda_i} \rho^{AB} \ket{\lambda_i}$. \\
It is well known that for an entangled system, unitaries acting on two different subsystems separately can be reduced to an identity acting on one subsystem and a modified unitary operating on the other subsystem\cite{Coecke_2010}. Considering this, we can express this nearest pure state in the following form:

\begin{align}
\ket{\psi}_\rho^{AB} = \frac{1}{\sqrt{2}}\left(\ket{H}^A\ket{\phi_H}^B + \ket{V}^A\ket{\phi_V}^B\right). \label{Eq:nearest_pure_state}
\end{align}
In this scenario, while performing coincidence measurements between Alice and Bob, if Alice has measured in $\{\ket{H}, \ket{V}\}$ basis, Bob's measurement in $\{\ket{\phi_H}, \ket{\phi_V}\}$ basis will result in a maximum correlation. From the nearest pure state, we find that the obtained $\ket{\phi_H}$ and $\ket{\phi_V}$ are nearly orthogonal, because of the very high Concurrence of $\ket{\psi}_\rho^{AB}$. So, as our modified measurement settings, we derive one of the orthogonal basis vectors $\ket{\phi_H}$ and find the angles of the waveplates to create a projection corresponding to this basis vector.

Similarly, if we consider the state in $\{\ket{D}, \ket{A}\}$ basis, the nearest pure state can be expressed as:
\begin{align}
\ket{\psi}_\rho^{AB} = \frac{1}{\sqrt{2}}\left(\ket{D}^A\ket{\phi_D}^B + \ket{A}^A\ket{\phi_A}^B\right), \label{Eq:nearest_pure_diagonal}
\end{align}
To obtain these modified bases we can do an inner product of the first qubit with the conventional basis with identity operated on the second. We then derive:
\begin{align}
    \ket{\phi_X}_{Unnormalized}{=}(\bra{X}\otimes I)\ket{\psi}_\rho^{AB}
    \label{Eq:nearest_pure_state_basis}
\end{align}
where $\ket{X}$ is the conventional basis vector.\\
After normalizing the modified basis vector we create the projection for measurement in the same basis. Experimentally to measure the photon state in a certain basis we use a combination of quarter-wave plate($QWP$), half-wave plate($HWP$) and polarisation beam splitter($PBS$). Jones matrices corresponding to $HWP$ and $QWP$ are given by \cite{hecht2017optics}:
\begin{align*}
    HWP=\begin{pmatrix}
        cos 2\alpha & sin 2\alpha\\
        sin 2\alpha & -cos 2\alpha
    \end{pmatrix}
\end{align*}
\begin{align*}
    QWP=\begin{pmatrix}
        cos^2\beta+isin^2\beta & (1-i)sin\beta cos\beta\\
        (1-i)sin\beta cos\beta & sin^2\beta+icos^2\beta
    \end{pmatrix}
\end{align*}
where $\alpha$ and $\beta$ are the angles between the fast axis of the waveplates and the horizontal plane of reference.
We know that a measurement at the transmitted port of the $PBS$, projects the state to $\ket{H}$. Using the newly obtained basis vector $\ket{\phi_X}$ information with the following equation, we can determine our new $HWP$ and $QWP$ angles i.e., $\alpha^{'}$ and $\beta^{'}$ corresponding to the projection for $\ket{\phi_X}$:
\begin{align}
    \ket{\phi_X}=QWP(\beta^{'})HWP(\alpha^{'})\ket{H}
    \label{projection}
\end{align}
As can be seen in Fig. \ref{ExptSchematic}, we have positioned detectors on both the transmitted and reflected ports of the $PBS$. Using Eq. \ref{projection} we find $\alpha^{'}$ and $\beta^{'}$ corresponding to projection $\ket{\phi_H}$ in the transmitted port at SPAD-B1. The same angles of the waveplates also create a projection of $\ket{\phi_V}$ in the reflected port of the $PBS$ at SPAD-B2. Similarly, we measure the photons in $\{\ket{\phi_D}, \ket{\phi_A}\}$ basis at detectors B3 and B4 respectively. By doing so, we counter any polarisation scrambling that may have happened because of the transmission through the quantum channel. Without any active feedback-based correction by choosing the correct MUB for measurement, we can now achieve secure communication protocols with low QBER and high Keyrate.\\
Our approach to polarisation correction is very different from any active feedback system. Active feedback involves actively manipulating the polarisation state of photons using external components such as wave plates, polarizers, and modulators. Another classical beacon signal on which polarisation measurements are performed to obtain the required feedback has to be employed in such a case \cite{Xavier_2009ActiveFeedbackwithBeacon,Li18ActiveFeedbackClassicalBeacon}. A few other ways of getting active feedback done, involve active polarisation tracking devices such as robotized polarisation correction based on an active control system\cite{RobotizedFeedback}, an active control system based polarisation tracking, polarisation basis tracking using the sifted key\cite{Toyoshima2011PolBasisTracking} etc. Hence, the system continuously monitors the polarisation state of the incoming beacon and applies corrective actions in real-time to maintain the desired polarisation state of the signal photons. Unlike Eq. \ref{Eq:nearest_pure_state_basis}, for active feedback, the unitary transformation has to be derived that caused the state change using another beacon source and its polarisation state, i.e. $U_{Tranformation}\ket{X}=\ket{\phi_X}$. With the unitary transformation information, active polarisation correction has to be employed using one of the above-discussed processes.\\
Passive feedback, on the other hand, does not involve real-time manipulation of the photon's polarisation state but relies on post-measurement correction. In our experimental scenario, we implement corrective actions by changing the measurement basis only after state tomography. So, the subsequent measurements for the protocol are done on a different measurement basis than the conventional ones. Instead of actively monitoring and manipulating the polarisation state of photons throughout the experiment, the correction for polarisation occurs passively by adjusting measurement settings, requiring minimal resources. 
\subsection{Stationary transmitter and receiver}
As a natural and more practically motivated follow-up to our MUB approach, we have been performing free space BBM92 protocols when Bob does his measurements using the new transformed bases obtained from the analysis discussed earlier. Our experimental configuration involves two separate facilities, each hosting specialized modules for Alice and Bob. These facilities are located at a distance of 50 meters from each other.
The process begins with the generation of entangled photon pairs in the same facility as Alice's equipment. These entangled photons are subsequently distributed to both Alice and Bob. Alice's photons are transmitted via a fibre optic connection, whereas Bob's photons undergo a more complex transmission process. They are first collected through a fibre optic link and then pass through a series of refractive optics implementing a telescopic arrangement. Finally, they traverse a 50-meter free-space link to reach Bob's location. With such an experimental setup we have performed QKD both in daytime and nighttime conditions.
\subsection{Daylight QKD}
In traditional QKD experiments, photons are sent over optical fibres or free space in a controlled laboratory environment with low levels of ambient light. However, extending QKD to the sender and receiver located at large distances from each other and to daytime scenarios presents unique challenges such as high background noise, atmospheric turbulence, etc. The QBER of the system increases significantly during the daytime, possibly much
higher than 11$\%$, if there is no adequate noise-filtering system.\\

\textbf{Background Noise:} Daytime environments are inherently brighter due to sunlight, which introduces a significant amount of background noise. This complicates the detection of the single photons, which serve as our quantum signals.\\

\textbf{Atmospheric Interference:} Earth's changing atmospheric conditions, lead to scattering and absorption of quantum signals, further reducing their fidelity and efficiency of being detected.\\

\textbf{Temperature Fluctuations:} Unlike our controlled laboratory temperature conditions, in an outside environment and particularly in the daytime, we record major temperature variations, which can affect the stability of any optical or optomechanical components, potentially degrading the performance of QKD systems.\\

Near future quantum networks will require daylight operation for continuous availability. This is because quantum networks are envisioned to be used for various applications, including secure communication, distributed quantum computing, and quantum sensing. For these applications to be successful, quantum networks need to be able to operate reliably, regardless of the time of day or weather conditions.\\

The Hughes experiment was the first to demonstrate the feasibility of daylight QKD over a distance of 1.6 Km in the year 2000\cite{buttler2000daylight}. A variety of noise reduction methods across the spectral, temporal, and spatial domains are employed to successfully remove noise photons from sunlight \cite{hughes2002practical,peloso2009daylight}. This comprehensive approach is necessary to reduce the QBER, which is a critical factor in ensuring the security and reliability of these systems. Some other approaches include the study on optimal wavelengths that are less affected by background noise from the sun\cite{avesani2021full,abasifard2023ideal,liao2017long}. The development of entanglement-based daylight QKD systems is still in its early stages, but it is a rapidly growing field of research. Peloso and team established the first daylight entanglement-based QKD continuously over several days
under varying light and weather conditions \cite{peloso2009daylight}. Recently, entanglement-based
QKD in daylight using a Quantum Dot-based photon source has also been implemented \cite{basset2023daylight}. \\



In the current experiment, we are employing spectral, temporal and spatial filtering together with the optimised measurement basis method for polarisation correction to achieve entanglement-based daylight QKD. 
The utilization of bandpass filters, offering a 10 nm FWHM transmission for spectral filtering, holds a crucial role in mitigating noise. Additionally, we evaluate the system's performance employing a 3 nm FWHM bandpass filter. Temporal filtering includes the choice of a coincidence window to minimize the chances of detecting background photons as valid signal photons. We can reject any photon pairs where the arrival times fall outside this window. This rejection effectively filters out most of the background noise, as it is unlikely to produce photon pairs with precise timing. A narrower window may provide better security but at the cost of a reduced key generation rate. Optimization algorithms that vary the coincidence window iteratively to obtain QBER under a specified limit and maximize the keyrate for such a scenario exist and are presented in our previous work. Further filtering and improving signal-to-noise ratio (SNR) can be done by employing the same optimization strategies. Spatial filtering is achieved by placing the pinhole apertures at the entrance of Bob's Setup and in front of the detectors. These apertures have dimensions that closely align with the diameter of the incident signal beam. The purpose of this configuration is to prevent unwanted single photons from entering the system. We observe that our MUB-based approach works very well even in daylight together with this filtering technique.\\

The daylight QKD experiment was performed on 16th, 20th and 22nd September 2023 from 08:00 AM to 18:00 PM (IST) with varying weather conditions and filters. The first QKD experiment was carried out on a sunny day with a 10 nm bandpass filter, followed by the second experiment on 20 Sept 2023 mild rainy day, which used a $3 nm$ bandpass filter. The third data collection occurred during sunny weather with a $10 nm$ Band Pass Filter. The aim was to analyze the results for keyrate and QBER throughout the day for different light conditions.  The data were taken with an interval of 2 hours.

\subsection{Night time QKD}
Night-time conditions often lead to lower background noise levels, creating a more favourable signal-to-noise ratio for QKD protocols.
It benefits from the relatively stable nighttime temperatures, levels of turbulence and atmospheric interference compared to daytime.\\

We will discuss 3 representative data sets for day and night QKD obtained on September 15th, 19th, and 21st, 2023 in different weather and filtering conditions. On September 15, 2023, the initial experiment occurred in ideal weather conditions, employing a $10 nm$ FWHM bandpass filter. The second dataset, collected on September 19, 2023, involved a switch to a $3 nm$ FWHM bandpass filter under clear skies. The third dataset, gathered on September 21, 2023, took place during rainy weather and utilized again a $10 nm$ bandpass filter. Our observations reveal that nighttime QKD yields a higher key rate and lower Quantum Bit Error Rate (QBER), as expected.

\section{Entangled Source}
\label{Entangled Source}
\begin{figure}
    \centering
    \includegraphics[trim=60 0 0 0,clip,scale=0.25]{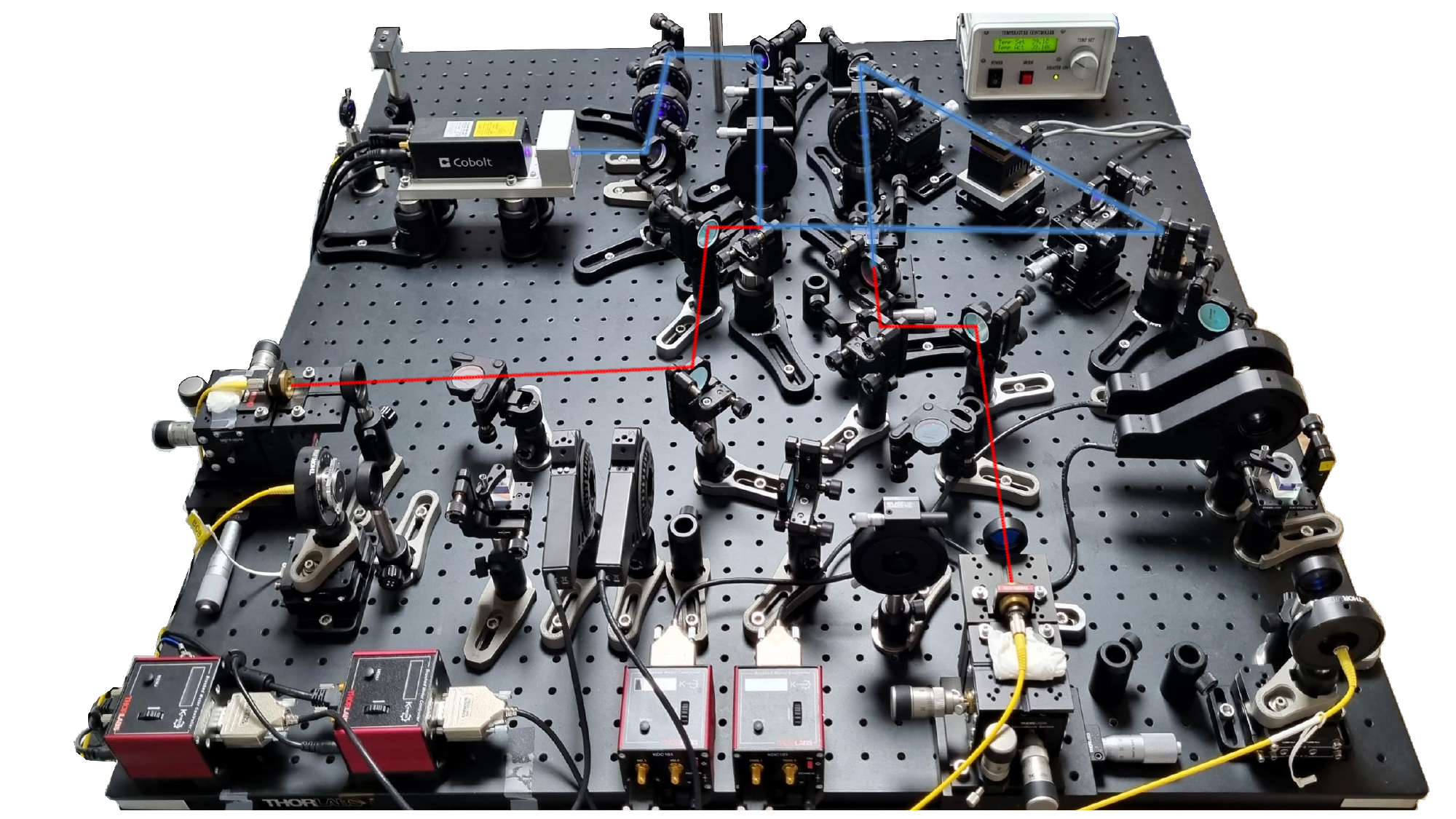}
    \caption{Type-II Entangled Photon source with PPKTP crystal under bidirectional pumping within a Sagnac Interferometer. The blue shaded line, a guide for the eye, illustrates the path of the pump beam both before and inside the interferometer. The red dashed line, also a visual guide, represents the trajectories of the entangled photon pair.}
    \label{fig:Source}
\end{figure}

Our experimental setup for implementing the BBM92 protocol involves a PEBS (Polarisation Entangled Bi-Photon Source). This source generates polarisation-entangled photon pairs through the Spontaneous Parametric Down-Conversion (SPDC) process. It uses a doubly pumped Type-II periodically-poled Potassium Titanyl Phosphate (PPKTP) crystal configured in a Sagnac arrangement (see Fig. \ref{fig:Source}). We use a set of waveplates to switch between bidirectional and unidirectional pumping of the crystal. Lenses are placed on either side of the crystal to focus and collimate the input and output beams respectively. As can be seen in Fig. \ref{fig:Source} we collect the entangled photons from the PBS's output ports and direct them into the single-mode fibres. We have the set-up for quantum state tomography also on the same optical table as the photon source. Our source yields polarisation-entangled single-photon pairs with a measured fidelity of $89\%$ with the Bell state $\ket{\Psi^+}_{12}=\frac{1}{\sqrt{2}}(\ket{H}_1\ket{V}_2+\ket{V}_1\ket{H}_2)$ and Concurrence of $0.9$.
We transmit these degenerate entangled single photons, operating at $810 nm$, to the Alice and Bob modules via two optical fibres. Each optical fibre is accompanied by a Fibre-Bench Polarisation Controller Kit from Thorlabs (PC-FFB-780). Utilizing these polarisation controllers, we can do synchronized tomography and obtain a maximum fidelity of $70\%$ between Alice and Bob located in facilities separated by 50 metres. On average, the source produces entangled photon pairs at a rate of $1 MHz$.\\

\section{Free-space BBM92 Experiment and Results}
\label{Experiment}



\subsection*{Experimental Set-up and Procedure}
\begin{figure*}
    \includegraphics[trim=30 140 40 60,clip,scale=0.5]{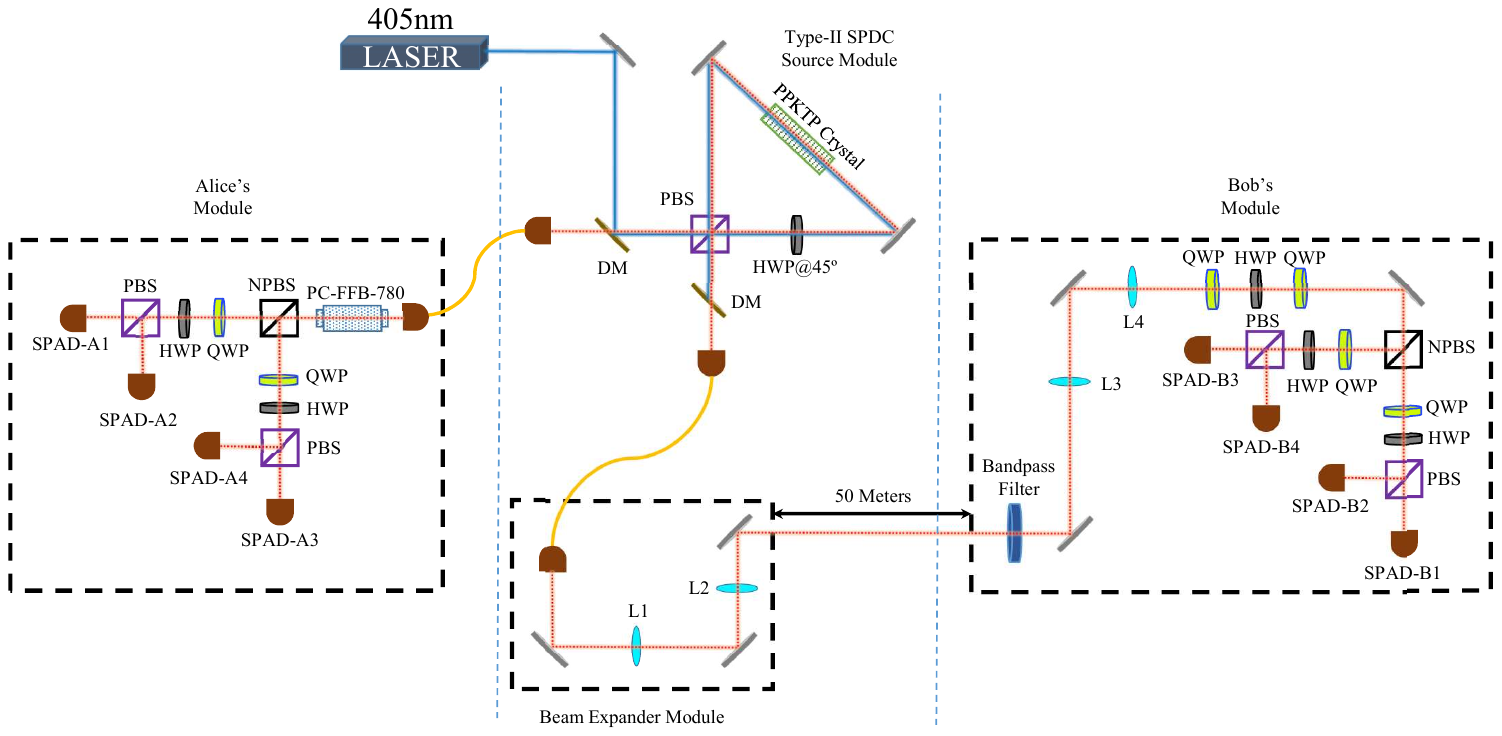}
    \caption{The figure illustrates the experimental configuration for a free-space quantum key distribution (QKD) protocol. The key components utilized and depicted in the schematic include a 405nm wavelength pump laser, a PPKTP (Periodically-Poled Potassium Titanyl Phosphate) crystal, optical elements such as a PBS (Polarizing Beam Splitter) and NPBS (Non-Polarizing Beam Splitter), QWP (Quarter Wave Plate), HWP (Half Wave Plate), DM (Dichroic Mirror), SPAD (Single Photon Avalanche Diode), dielectric mirrors, lenses, and various electronic components.} 
    \label{ExptSchematic}
\end{figure*}

\begin{figure*}
    \centering
    \includegraphics[scale=0.34]{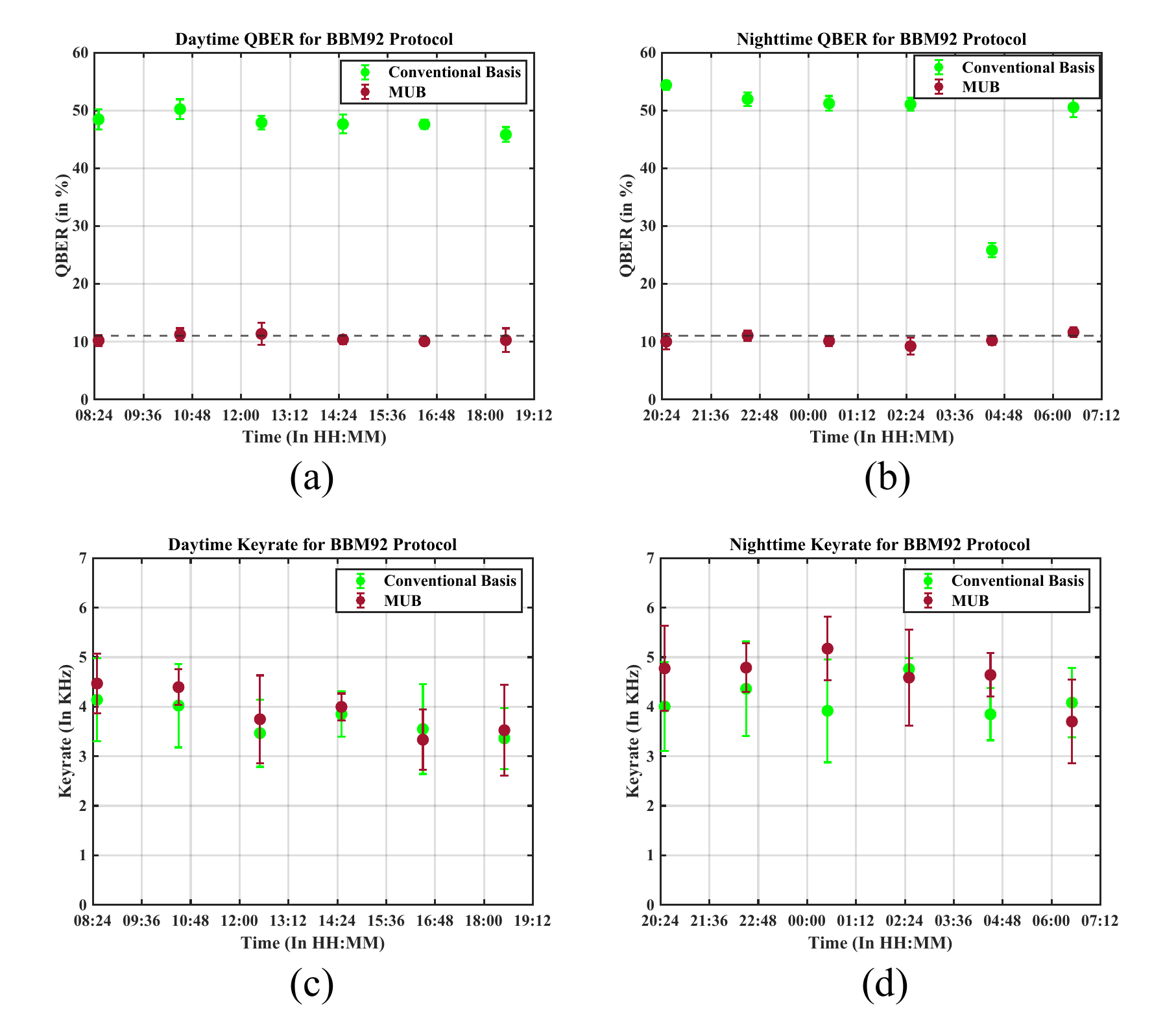}
    \caption{Experimental free-space QKD data for September 15th-16th. Fig. {(a) and (c)} compare the QBER and the Keyrate obtained for protocols carried out in the daytime with the conventional basis vs. the corrected basis. Fig. {(b) and (d)} compare the QBER and the Keyrate obtained for protocols carried out in the nighttime with the conventional basis vs. the corrected basis. Average daylight keyrate and QBER = ($3.9118\pm0.7339 KHz$ and $10.5518\pm1.3428\%$) respectively. Average nighttime Keyrate and QBER = ($4.6118\pm0.8088 KHz$ and $10.3545\pm1.2501\%$) respectively.}
    \label{fig:Data-15-16}
\end{figure*}
\begin{figure*}[t]
    \centering
    \includegraphics[scale=0.34]{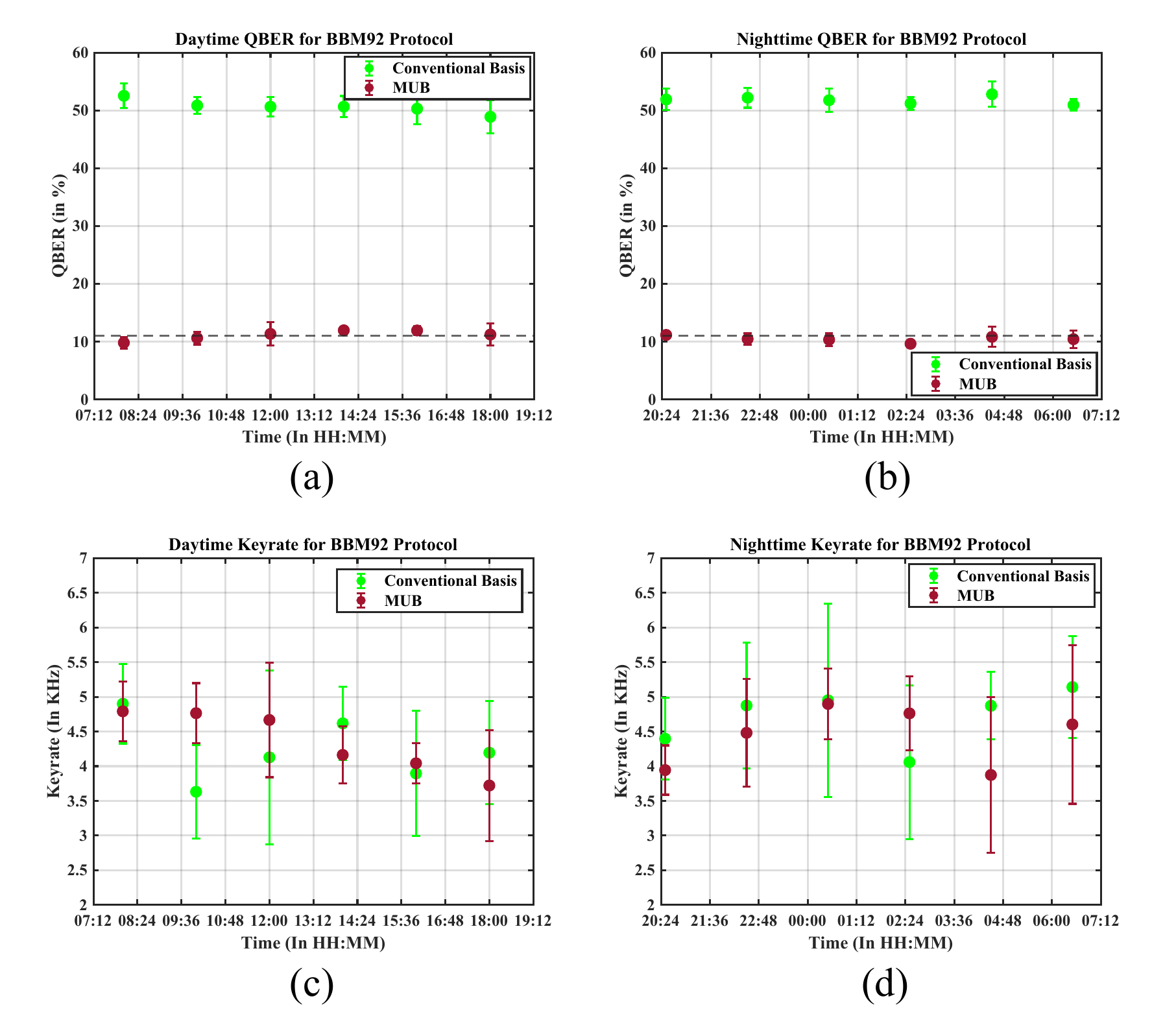}
    \caption{Experimental free-space QKD data for September 19th-20th. Fig. {(a) and (c)} compare the QBER and the Keyrate obtained for protocols carried out in the daytime with the conventional basis vs. the corrected basis. Fig. {(b) and (d)} compare the QBER and the Keyrate obtained for protocols carried out in the nighttime with the conventional basis vs. the corrected basis. Average daylight keyrate and QBER = ($4.3578\pm0.6626 KHz$ and $11.1301\pm1.4384\%$) respectively. Average nighttime Keyrate and QBER = ($4.4267\pm0.8293 KHz$ and $10.4533\pm1.1719\%$) respectively.}
    \label{fig:Data-19-20}
\end{figure*}

\begin{figure*}
    \centering
    \includegraphics[scale=0.34]{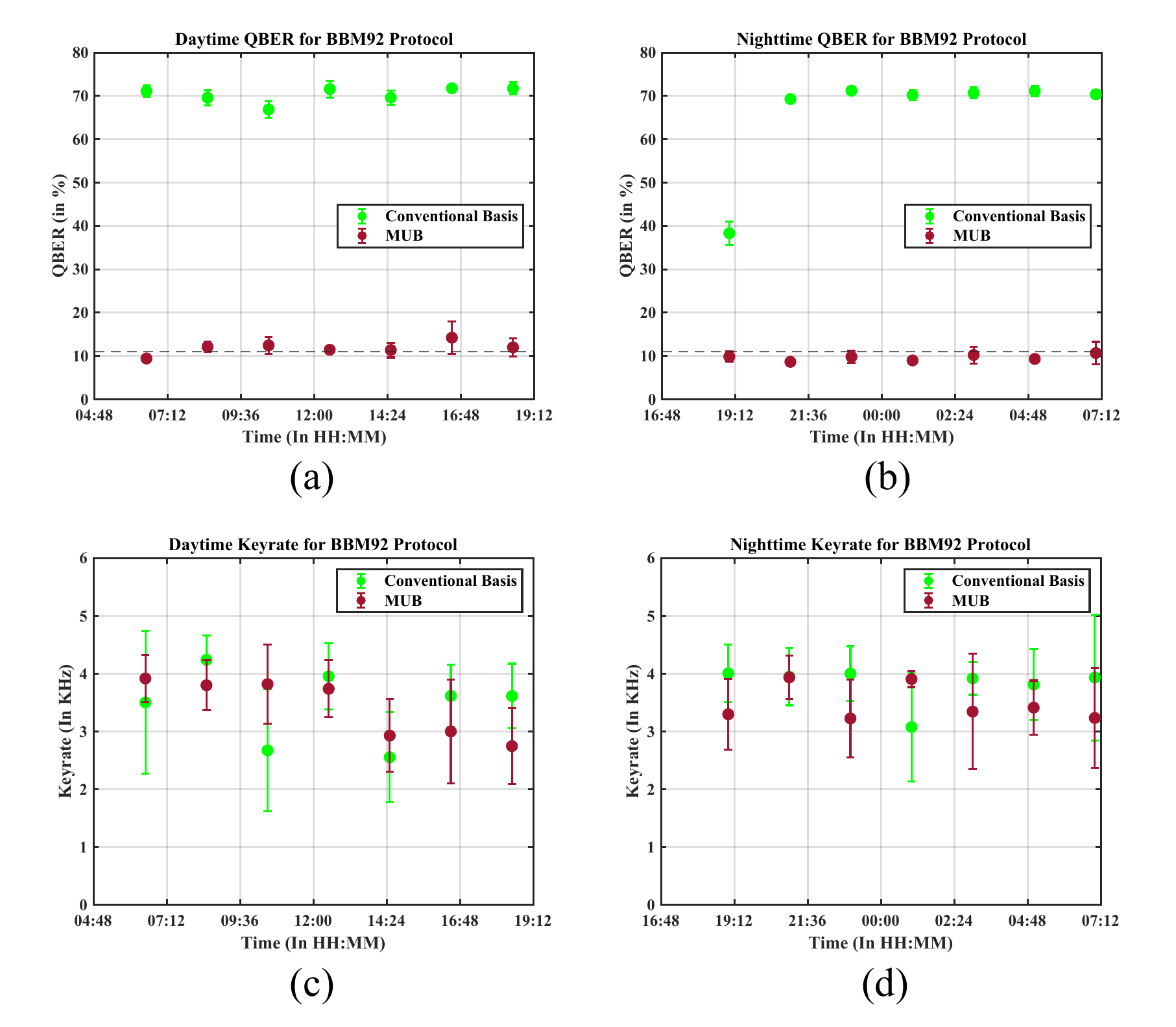}
    \caption{Experimental free-space QKD data for September 21st-22nd. Fig. {(a) and (c)} compare the QBER and the Keyrate obtained for protocols carried out in the daytime with the conventional basis vs. the corrected basis. Fig. {(b) and (d)} compare the QBER and the Keyrate obtained for protocols carried out in the nighttime with the conventional basis vs. the corrected basis. Average daylight keyrate and QBER = ($3.4223\pm0.7364 KHz$ and $11.8580\pm2.2684\%$) respectively. Average nighttime Keyrate and QBER = ($3.4801\pm0.6574 KHz$ and $9.6413\pm1.4710\%$) respectively.}
    \label{fig:Data-21-22}
\end{figure*}
\textbf{Transmitter:} The transmitter module includes the entangled photon source, two $5m$ single-mode fibres (SMF), and a refractor telescope arrangement for beam expansion between the source and Bob to achieve the free space link. The pair of entangled photons is distributed between Alice and Bob.
Photon 1 goes to Alice's detection module, guided by the SMF, whereas Photon 2 is guided through another SMF and evolves into the telescopic arrangement. This is where we perform beam expansion using a combination of two lenses to have $4X$ magnification. This counters the beam divergence because of the free space transmission. At the initial alignment stage, we use $633 nm$ and $810 nm$ continuous-wave lasers to establish the link and verify the magnification after the lens combination.\\

\textbf{Receiver:} Alice, as one of the receivers, has the optics aligned for the state tomography and QKD protocol experiments. Output photons from the SMF go through the polarisation controller, followed by a combination of QWPs, HWPs, and PBSs, as shown in Fig. \ref{ExptSchematic}. These optics help in creating different projections to measure the single photons. There are four different single-photon detectors at the end.\\
Similarly, Bob has the same polarisation optics arrangement at its end, followed by four more single photon detectors. Apart from this, Bob's module also has a $5X$ demagnification setup consisting of two lenses and a polarisation controller setup with two QWPs and one HWP for the initial polarisation correction (see Fig.\ref{ExptSchematic}).\\

\textbf{Time Synchronization:} Synchronization with GPS ensures that Alice and Bob's quantum systems are properly aligned in time, minimizing the risk of attacks or information leakage due to timing discrepancies. It's an essential component of QKD systems to achieve secure key exchange over long distances. Alice and Bob use highly accurate GPS clocks to establish a common time reference. PPS (Pulse Per Signal)  signals provide a reference point for the synchronisation clock. These clocks are synchronized to the signals received from multiple GPS satellites, providing a very accurate measure of time. With synchronized timing, Alice and Bob can perform the QKD protocol securely. They can now share their measurement outcomes, typically using a classical communication channel, to establish a secret cryptographic key that is secure against any eavesdropping attempts by Eve. We use time tagger TT20 and TTULTRA from Swabian instruments at Alice and Bob's ends, respectively, to record the timestamps.\\

Our initial alignment and optimization for the experiment involve the following steps: We establish the link between two laboratories using a visible $633 nm$ laser. As our single photons are of $810 nm$, we use a laser of the same wavelength to verify magnification and collimation. Also, the optimization of coupling with all four couplers at Bob's end is done using the same laser. As we switch to single photons, we perform an initial polarisation correction using the polarizer controllers at both Alice and Bob end with our PPKTP crystal inside the Sagnac interferometer pumped in one direction. With these initial tests, we consider the free space link ready to be used as our quantum channel. Then, with bidirectional pumping of the PPKTP crystal, we generate and transmit entangled photon pairs to Alice and Bob, and we conduct quantum state tomography, thoroughly characterizing the initial quantum states involved in our experiment. On every incoming photon, both parties measure the polarisation in $H/V$, $D/A$, and $R/L$ bases to generate the time stamps on which the cross-correlation is performed to detect the (anti-)correlations. For a complete tomography, measurement is done in $36$ different projections.
\begin{figure*}
    \centering
    \includegraphics[scale=0.5]{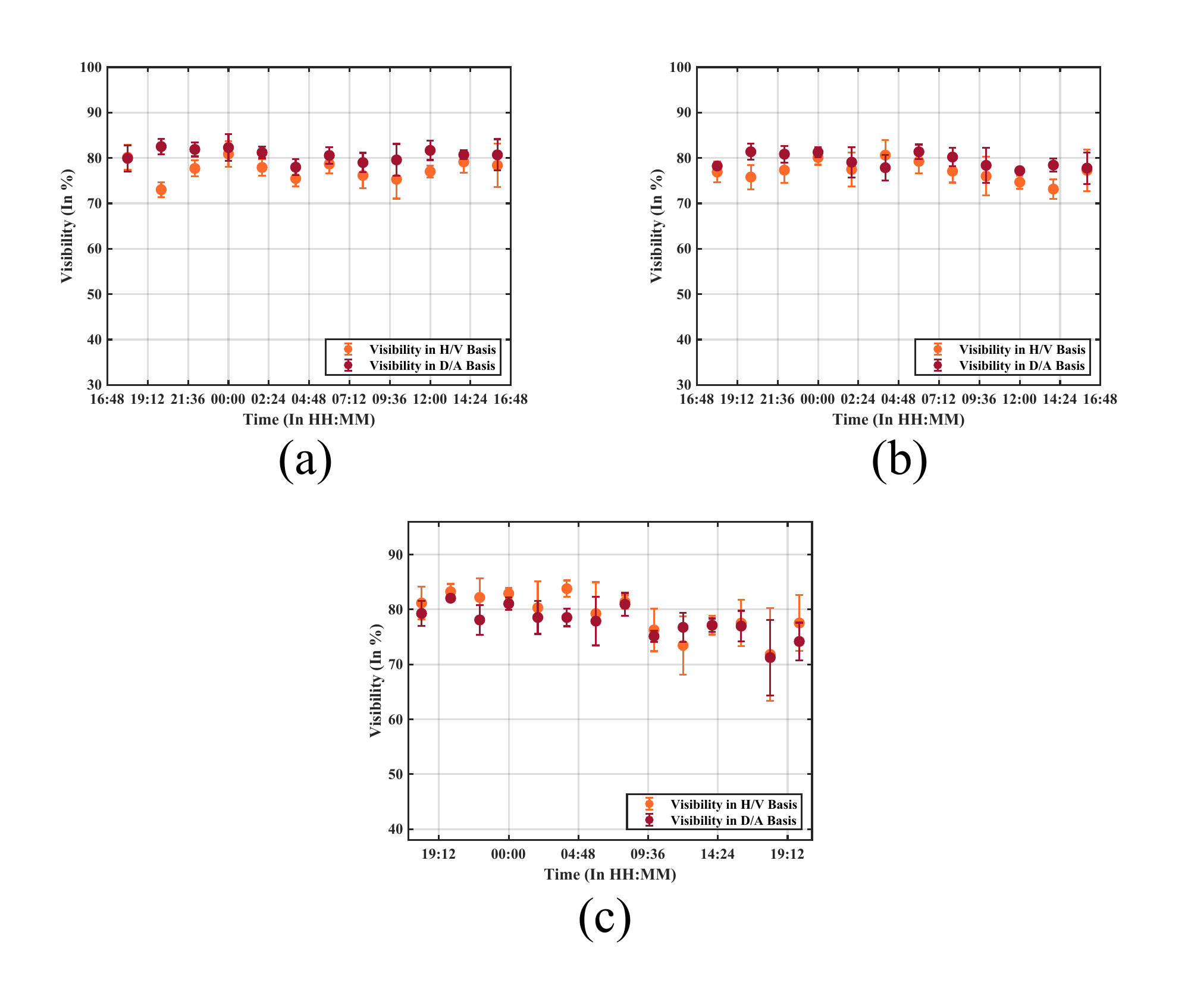}
    \caption{Visibilty in H/V and D/A basis. Fig. (a), (b) and (c) correspond to data taken on the 15th-16th, 19th-20th and 21st-22nd Sep respectively.}
    \label{Visibility}
\end{figure*}
For the QKD protocol, using the nearest pure state analysis, we determine the new MUBs. Once identified, we adjust the waveplates within Bob's module accordingly to align with the projection associated with these new MUBs i.e. $\phi_H/\phi_V$ and $\phi_D/\phi_A$. Meanwhile, at Alice's station, we perform photon measurements using the conventional $H/V$ and $D/A$ bases. The choice of these MUBs results in better QBER and keyrate than the conventional basis choice at both Alice's and Bob's end. The signal-to-noise ratio is consistently higher for the corrected basis. The visibility in each respective basis served as an indicator of these characteristics. Throughout our experimental runs, both the $H/V$ and the $D/A$ basis exhibited very high and comparable visibility with each other (see Fig. \ref{Visibility}). Therefore, for all runs of the protocol, we have chosen these two bases to measure the photons in.\\
For QKD protocol data acquisition, with our SPAD combinations, we do coincidence measurements in $8$ different projections, where we expect $4$ projections to give us high coincidence contributing to the actual signal and $4$ others to give low coincidence and represent the noise in the system. The sum of all coincidence detections divided by the protocol runtime will give us the Keyrate (see Eq.\ref{Keyrate}). Meanwhile, the ratio of $4$ unwanted coincidence measurements (representing noise) and the sum of all coincidences will be the error in the key distribution. Again this error divided by the protocol runtime will provide us with the QBER (see Eq.\ref{QBER}).
\begin{equation} 
\resizebox{1\columnwidth}{!}{$Keyrate=\frac{C(A1,B1)+C(A1,B2)+C(A2,B1)+C(A2,B2)+C(A3,B3)+C(A3,B4)+C(A4,B3)+C(A4,B4)}{T}(bps)$}
\label{Keyrate}
\end{equation}

\begin{equation}
    \resizebox{1\columnwidth}{!}{$QBER=\frac{[C(A1,B2)+C(A2,B1)+C(A3,B4)+C(A4,B3)]*100}{T*[C(A1,B1)+C(A1,B2)+C(A2,B1)+C(A2,B2)+C(A3,B3)+C(A3,B4)+C(A4,B3)+C(A4,B4)]}$}
    \label{QBER}
\end{equation}
Here, A1, A2, A3, A4 and B1, B2, B3, B4 represent detectors at Alice's end and at Bob's end respectively (see Fig. \ref{ExptSchematic}).\\
T represents the Acquisition time. We have chosen T to be 10secs for all our protocols.\\ 
Our objective is to identify coincidence peaks between various detectors located at Alice and Bob's ends. This involves the precise adjustment of timing delays and establishment of a suitable coincidence window. Through this process, we derive parameters such as Keyrate and QBER, which quantify the level of errors in the quantum communication process.

\section{Results}
\label{Results}
We performed the free-space QKD experiments on different days under changing weather conditions. In this section, we present data from a series of QKD experiments conducted on different days, each spanning a continuous 24-hour data collection period, including both daytime and nighttime QKD sessions. 

\begin{figure}
    \centering
    \includegraphics[trim=50 20 20 20,clip,scale=0.5]{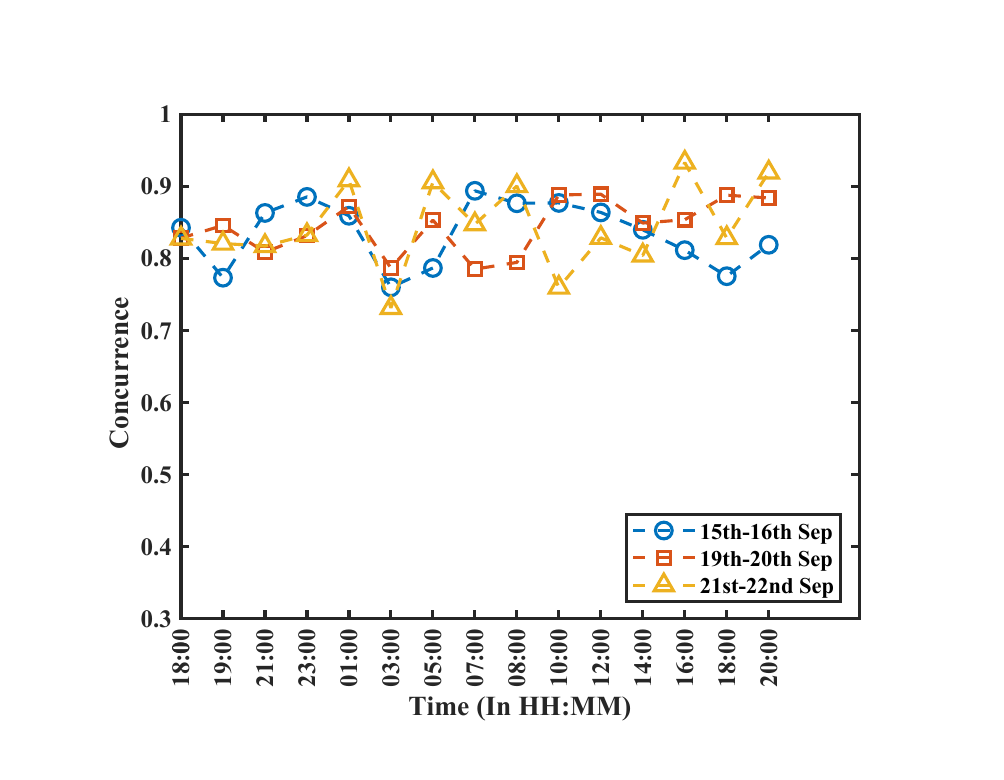}
    \caption{Varying Concurrence over 24hrs on different days of the experiment}
    \label{fig:Concurrence}
\end{figure}

\begin{figure}
    \centering
    \includegraphics[trim=50 20 20 20,clip,scale=0.5]{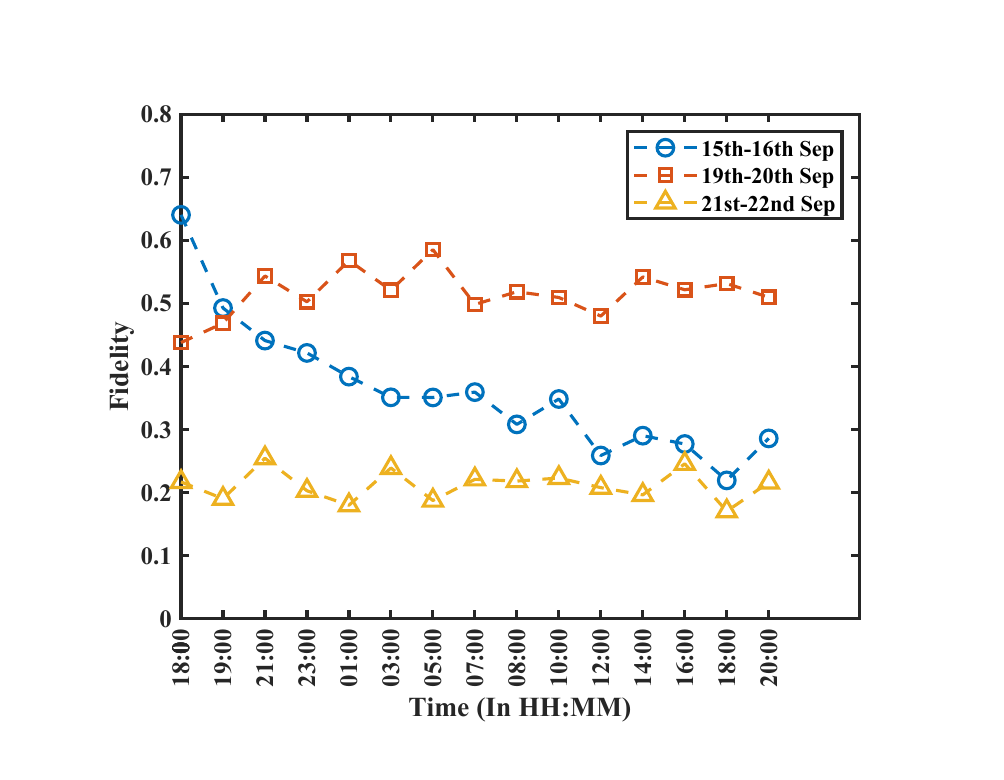}
    \caption{Varying fidelity over 24hrs on different days of the experiment}
    \label{fig:Fidelity}
\end{figure}

Specifically, we focus on three key data sets, results from which are given in Fig. \ref{fig:Data-15-16}, \ref{fig:Data-19-20} and \ref{fig:Data-21-22} respectively.
On the night of September 15th-16th 2023, our experiment took place under good weather conditions with a clear sky, providing ideal conditions for our QKD setup. During the daytime on 16th, we had a sunny day. As illustrated in Fig. \ref{fig:Data-15-16}, the data obtained on this day demonstrates good performance, characterized by a low QBER and a high keyrate. We obtained the average daylight keyrate and QBER of ($3.9118\pm0.7339 KHz$ and $10.5518\pm1.3428\%$) respectively and night time Keyrate and QBER of ($4.6118\pm0.8088 KHz$ and $10.3545\pm1.2501\%$) respectively.\\
Let us now discuss the results of the data that we took on September 19th-20th 2023. We conducted our experiments in the presence of partially rainy conditions. Moreover, for this data collection, we employed a narrower $3nm$ Full-Width at Half Maximum (FWHM) bandpass filter, unlike the $10nm$ bandpass filter used on other days to understand the effect of further narrower spectral filtering on the experiment. This specific spectral filtering change as well as the change in ambient weather conditions resulted in a keyrate that is quite comparable to the previous data, as shown in Fig. \ref{fig:Data-19-20}. We concluded that our system works the same with different spectral filtering at the receiver's end. This is also because of the filtering that is employed at the source itself. Hence, our single photon numbers that are getting transmitted through the free space don't get much affected because of the narrower bandpass filter at the receiver end. For this data we measured the average daylight keyrate and QBER of ($4.3578\pm0.6626 KHz$ and $11.1301\pm1.4384\%$) respectively and night time Keyrate and QBER of ($4.4267\pm0.8293 KHz$ and $10.4533\pm1.1719\%$) respectively.\\
Lastly, on September 21st-22nd 2023, our data collection took place under a mix of weather conditions: light rain during the night followed by clear daytime skies. We obtained the average daylight keyrate and QBER of ($3.4223\pm0.7364 KHz$ and $11.8580\pm2.2684\%$) respectively and night time Keyrate and QBER of ($3.4801\pm0.6574 KHz$ and $9.6413\pm1.4710\%$) respectively. Despite the variability in weather conditions, our results consistently reveal robust performance, with both good Keyrate and QBER that generally maintain information-theoretic security. Fig. \ref{fig:Data-21-22} has the representative plots for the final dataset. Also, it is worth noting that on the last day, we had a fidelity of $~0.2$ with the Bell state $\ket{\Psi^+}$ as shown in Fig. \ref{fig:Fidelity}. This is reflected in the data corresponding to the QBER ($>70\%$) for the conventional basis. Regardless of the lower fidelity, we still get generally $<11\%$ QBER for our protocols with the correct MUB in the nighttime. At the same time, we had comparatively higher fidelity and Concurrence (see Fig. \ref{fig:Concurrence}) on the other two days resulting in lower QBER with the conventional basis. MUB approach for passive correction holds for these different conditions of state preparation as well as dynamic changes in the environment. The results presented here can be seen to have mean and error bars associated with QBER being above the information-theoretic security threshold at times. All the data represents QBER and Keyrate extracted from raw timestamps. Though we are generally maintaining information-theoretic secure QBER, some local fluctuation leads to higher QBER in some instances. Further optimization in SNR will be necessary to mitigate such situations. From our earlier work\cite{PatentQuIC} we have optimization algorithms that can limit the QBER within $11\%$ and optimize Keyrate and Key-symmetry. We would investigate the application of  further optimization analysis to our current results at a later stage.\\
Throughout all the presented plots, we compared the data obtained using the conventional measurement basis to our MUB approach. The consistent trend we observed is that selecting the corrected MUB as the measurement basis produces appropriate results, while the conventional measurement basis yields a very high QBER.
These results validate the adaptability and robustness of our MUB approach, showcasing its effectiveness even in varying environmental conditions. This ensures the efficiency of our passive correction mechanism in free space, particularly over extended distances.\\

\section{Discussion}
\label{Discussion}
In this work, we have tested the practicality of our recently introduced MUB-based passive correction mechanism against the polarisation scrambling that happens due to both fibre birefringence as well as the difference in frame of reference of sender and receiver, by testing the efficacy of the approach over entanglement based QKD established over a free space atmospheric link between two buildings in our institute. Our results show promise in achieving QBER that is below the information-theoretic security threshold and a consistently high keyrate in varying light as well as weather conditions.\\

Our current experiment focuses on stationary source and receiver platforms. However, real-world long-distance applications often involve one or both platforms in motion. Previous works have explored such scenarios with terrestrial vehicles\cite{kwiat2022autonomous}, drones\cite{carrasco2011low}, aeroplanes\cite{nauerth2013air}, hot air balloons\cite{wang2013direct}, and, for very long-distance communication, satellites\cite{liao2017satellite} as carriers for either the source or receiver modules. A promising future research investigation could be towards testing our measurement-based correction mechanism in contexts with moving sender or receiver.\\ 

A rate-limiting step for our current methodology is the need to perform quantum state tomography before the start of the QKD protocol. Additionally, our method utilizes single photons in quantum state tomography, which are valuable but finite resources. While we save resources compared to active polarization feedback, some photons are inherently lost in passive feedback. Future research could involve a resource budget analysis, comparing active and passive feedback in practical real-life scenarios. Another promising future research direction would be towards investigating and establishing faster techniques of state analysis as well as employing variants of algorithms\cite{qkdSim} to post-process the data to further optimise the signal-to-noise ratio and leading to higher key rates and reduced QBER, ultimately improving the efficiency and performance of quantum key distribution protocols.

\section*{Data availability}

The data supporting the findings of this study may be available from the corresponding author upon reasonable request.

\section*{Code availability}

The codes supporting the findings of this study may be available from the corresponding author upon reasonable request.

\section*{Acknowledgments}

 US would like to thank the Indian Space Research Organization for support through the QuEST-ISRO research grant. We thank S. Chatterjee, K. Goswami, S. Prabhakaran and S. Sujatha for their initial technical assistance.

\section*{Author contribution}
SRB and MG contributed equally to the work. SRB and MG performed the experiments and data analysis under US's supervision. US conceived of and supervised the experiment. 

\section*{Competing interests}

The authors declare no competing interests.

\bibliographystyle{naturemag}
\bibliography{Reference.bib}

\end{document}